# Current Trends in the Use of Eye Tracking in Mathematics Education Research: A PME Survey

**Achim J. Lilienthal**

Örebro University, Sweden

**Maike Schindler**

University of Cologne, Germany

*Eye tracking (ET) is a research method that receives growing interest in mathematics education research (MER). This paper aims to give a literature overview, specifically focusing on the evolution of interest in this technology, ET equipment, and analysis methods used in mathematics education. To capture the current state, we focus on papers published in the proceedings of PME, one of the primary conferences dedicated to MER, of the last ten years. We identify trends in interest, methodology, and methods of analysis that are used in the community, and discuss possible future developments.*

## 1 Introduction

The resolution of the human eye provides a high-resolution image of the environment only in a small region, called the fovea centralis. The eyes must scan over areas in the environment about which detailed information is needed (Henderson, 2003; Rayner, 1998). Conversely, the sequence of eye movements, which brings areas of interest into foveal vision, allows drawing conclusions about where the attention of a person was directed to and what cognitive processes may have caused the observed gaze pattern.

The actual interpretation of gaze patterns, however, is non-trivial. Often, it is assumed that the eye-mind hypothesis (Just & Carpenter, 1980, 1976) holds, which claims that the eyes hover for a prolonged time (typically 100ms to 600ms) over those areas the mind is currently processing. However, a person may also gaze at an area without registering it and objects can also be recognized with peripheral instead of foveal vision (Holmqvist et al., 2011). Mapping gaze patterns to cognitive processes is thus afflicted with uncertainties and in general not bijective (Schindler & Lilienthal, 2017, 2019). If a person fixates an area, this could indicate, for example, difficulties to extract information from that area (Jacob & Karn, 2003), mental calculation (Hartmann et al., 2015), or boredom.

Despite these difficulties, the possibility to study patterns of attention and infer cognitive processes is appealing, especially to MER. Learning situations and mathematical tasks (the stimuli) can be designed to reduce uncertainty—and a domain-specific interpretation can also reduce ambiguity (Schindler & Lilienthal, 2019). Hence, the number of publications that involve the analysis of gaze patterns in educational research and MER in particular is growing in recent years (ibid.; Was, Sansosti, & Morris, 2017).

ET devices record movements of the pupil and identify gaze points by back-projecting from the fovea to the surrounding scene, thus gathering information about a person's visual attention. The interest in ET in MER is fueled by the availability of increasingly affordable, advanced, and accurate ET technology. The possibility to use inexpensive, massive computational resources makes ET in combination with a partially automated analysis also interesting for mathematics education practitioners.

PME, one of the primary conferences dedicated to MER, reflects the trend towards increasing interest in ET. Being at the forefront of MER, the PME conference hosted, e.g., two workshops about ET. First, at PME-38 (2014 in Vancouver, Canada) a Working Session about the "The use of eye-tracking technology in MER" (Barmby et al., 2014, p. 253) introduced the topic with the aim to "discuss the potential of this innovative approach [eye-tracking technology] to mathematics education research, including ways in which the approach can be superior to other methodological approaches" (ibid.) and to clarify "how eye-tracking technology can be incorporated into quantitative and qualitative approaches" (ibid.). A more recent Working Group at PME-42 (2018 in Umeå, Sweden) about "Eye-tracking in MER: A follow-up on opportunities and challenges" (Schindler et al., 2018, p. 209) took up the intensive hardware development since the Working Session at PME-38 and set out to discuss the implications of "on-line ET by ordinary web cameras [that may] become a part of everyday e-learning user experience in the next few years" (ibid.) and "challenges regarding setups of studies and methods of analysis" (ibid.), especially in connection to other recent developments such as "dual and multiple ET: the analysis of two or more persons' eye movements that allows for studying collaboration and the teaching/learning process in vivo" (ibid.).

The mentioned workshops at PME-38/42 demonstrate the interest in ET in MER. This interest is driven by recent hardware and methodological developments as well as new challenges and opportunities. In this situation, we see the need to reflect on the current status of the literature and to analyze trends and possible future directions as indicated by the ET related papers published in the PME proceedings. The PME peer-review process assures high quality. At the same time conference proceedings are more up-to-date compared to journal publications with longer publication processes, which is important for quickly developing research topics. For this paper, we thus analyzed all publications in the PME proceedings of the last ten years that mention ET.

A previous overview published at PME-39 as description of a Research Forum focused on ET for one particular purpose (analysis of strategy use in mathematical tasks) and on one particular class of ET devices. Instead, our aim is to analyze, categorize, and summarize all ET related papers at PME. Based on the analysis of the PME proceedings of the last ten years we ask the following research questions:

*RQ1: How did the number of ET related papers evolve over time?*

Different technologies are available to track eye movements. It is relatively clear that video-based systems, which track one or both pupils with one or more cameras, are mostly used in MER. Two major categories of these cameras are available: portable eye-trackers (ET glasses) or remote eye trackers that are attached to a screen which shows the visual stimuli (Holmqvist et al., 2011). Correspondingly, we ask:

*RQ2: What ET equipment is used in the MER community?*

The interpretation of ET data is non-trivial. It often involves initial computation of features that characterize the raw data, such as fixations (periods typically between 100ms to 600ms, in which the eyes hover over a small area), interjacent saccades (very fast movements between fixations with a duration of 20ms to 40ms), and dwell times (length of sequences of fixations on a particular area of interest); or a laborious manual coding. The actual interpretation of these features or gaze-overlaid videos is one of the key issues in most ET studies due to the mentioned uncertainties and non-bijectivity of gaze patterns. Our third research question is thus:

*RQ3: What analysis and (result) presentation methods are used in MER?*

## 2 Methodology

### 2.1 Full keyword search

We first identified all relevant papers through a *full text key word search* in the proceedings of PME-34 until PME-42. We systematically searched the electronic version of the proceedings using the following (not case-sensitive) search terms "Eye Tracking" and "Eye-Tracking", "Gaze Tracking" and "Gaze-Tracking", "Eye Movement" and "Eye-Movement", "Gaze Movement" and "Gaze-Movement". Figure 1 (top) shows how often the search terms appeared in total depending on the year. In particular, one can see an overall increasing trend and that before 2012 the search terms hardly occur at all.

### 2.2 Analysis and coding of articles

We then read all articles found and identified those papers that actually use ET instead of only, e.g., referring to ET-related papers or mentioning plans to carry out ET research. In this step we also coded the papers with respect to our research questions.

## 3 Results

### 3.1 RQ1: How did the number of ET related papers evolve over time?

The evolution of the number of contributions published in the PME proceedings, which relate to ET is shown in Figure 1 (bottom). We found 33 contributions in total, none before 2013 (PME35). This includes 13 research reports, 12 oral (or short oral) communications, 4 posters, 2 plenary lecture contributions and the descriptions of the two working groups mentioned before in this paper. It is not possible to identify a clearly rising trend. So far most ET contributions were published at PME40 but this could be an outlier. However, it is remarkable, that the highest number of research reports (5, i.e., 38.5% of all) were accepted at the most recent PME (PME42).

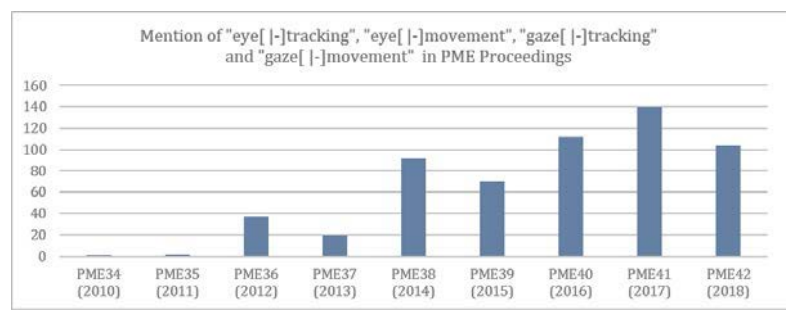

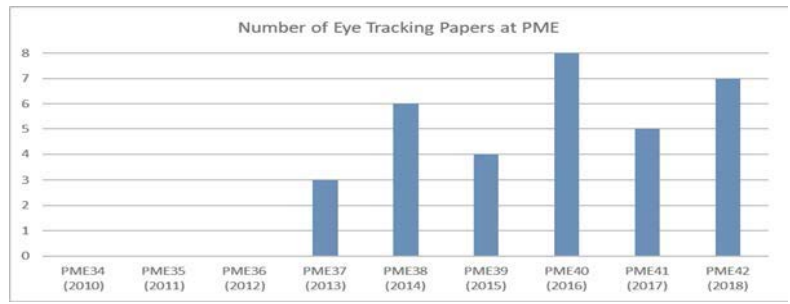

**Figure 1. (top) Number of appearances of ET related phrases. (bottom) Number of published PME contributions related to ET**

## 3.2 RQ2: What ET equipment is used in the MER community?

The majority of papers (21) use or discuss methodology related to remote eye trackers (Beitlich et al., 2014; Beitlich et al., 2015; Beitlich & Obersteiner, 2015; Chumachenko, 2014; Chumachenko et al., 2014; Chumachenko & Shvarts, 2016; Garcia Moreno-Esteva, 2016; Inglis & Alcock, 2018; Lehner et al., 2016; Lin & Ufer, 2016; Lithner, 2015; Obersteiner et al., 2014; Shvarts, 2013; Shvarts, 2014; Shvarts & Cumachenko, 2013; Shvarts & Zagorianakos, 2016; Strohmaier et al., 2016; Strohmaier et al., 2018; Tsai et al., 2014; Wu, 2015; Wu, 2017). A trend that seems to gain momentum more recently is towards portable ET glasses, which are used in 11 papers in total with 8 of them (72.7%) in the last two PMEs (Dewolf et al., 2013; Haataja et al., 2018; Hannula et al., 2017; Hannula, 2018; Hannula & Williams, 2016; Lilienthal & Schindler, 2017; Schindler & Lilienthal, 2017; Schindler et al., 2016; Schindler & Lilienthal, 2018; Shvarts, 2017; Shvarts, 2018). Almost all ET devices used are commercial systems. The notable exception are the ET glasses developed at the University of Helsinki (Toivanen, Lukander & Puolamäki, 2017). Another possible current trend is towards synchronized systems of two and more ET devices that allow to study interactions and social learning situations, both with portable eye trackers (Haataja et al., 2018; Hannula et al., 2017; Hannula, 2018; Lilienthal & Schindler, 2017; Shvarts, 2018) and remote eye trackers (Shvarts & Zagorianakos, 2016).

## 3.3 RQ3: What analysis and (result) presentation methods are used in MER?

We found in total 26 ET studies (from here on we omit referring to the PME papers; they are all cited above) for which we could determine the way the analysis was carried out. To our surprise, this was not possible for a small number of papers. The analysis methods fall into three different categories. Most often (15 times), quantitative analyses of features derived from the ET data were carried out. These features include the number or duration of fixations, dwell times, or the number or directions of saccades. In 9 cases, a manual (and therefore very laborious), typically qualitative analysis of ET videos was used for analysis. In one of these 9 cases, gaze paths were analyzed instead of the full gaze-overlaid video. The six methodological contributions we found in the PME proceedings address mostly dual and multiple ET and there is one paper that is dedicated to the automatic analysis of ET data for the predication of problem-solving success (Garcia Moreno-Esteva et al., 2016).

In addition to standard presentation of the results using statistical tests (primarily based on quantitative analysis of features derived from the ET data) and reports of the analysis of gaze-overlaid videos (partially supported by screenshots or linked videos), researchers also developed new ways of presenting their results. For instance, they show the mathematics related interpretation directly through paragraph plots (Inglis & Alcock, 2018) or the gaze synchrony graph (Hannula, 2018).

## 4 Discussion and Outlook

Our analysis of the ET related papers that were published in the PME proceedings over the last ten years showed that interest in ET is evident for the last five years. It also seems as if the number of papers and the numbers of ET being mentioned in the PME proceedings are increasing and more ET related papers are accepted as research reports. We also see a trend towards the use of portable ET equipment (ET glasses) and dual/multiple ET. Finally, we identified three major categories of analysis methods.

We believe that these trends are likely to continue as it seems the mathematics education community only starts to exploit the potential of ET. We further believe that methods of automatized evaluation will become more popular as they can help researchers to cope with the enormous effort to work with ET data and promise to be transferrable into the realm of education practitioners. Finally, we note that the potential of web-camera based ET, which may become omnipresent, is currently under-exploited and may become a popular research.